\documentclass[amsmath,amssymb,twocolumn,pre,floatfix]{revtex4}
\usepackage{graphicx}
\usepackage{epstopdf}
\def\D{{\mathrm d}}
\def\E{{\mathrm e}}
\def\I{{\mathrm i}}
\newcommand{\tf}[1]{\textstyle\frac #1}
\newcommand{\bs}[1]{\boldsymbol #1}
\begin{document}
\title{Vector Nematodynamics with Symmetry-driven Energy Exchange }
\author{L.~M. Pismen}
\affiliation{Department of Chemical Engineering, Technion -- Israel Institute of Technology, Haifa 32000, Israel}

\begin{abstract}
We review inadequacy of existing nematodynamic theories and suggest a novel way of establishing relations between nematic orientation and flow based on the \emph{local} symmetry between simultaneous rotation of nematic alignment and flow, which establishes energy exchange between the the two without reducing the problem to near-equilibrium conditions and invoking Onsager's relations. This approach, applied in the framework of the vector-based theory with a variable modulus, involves antisymmetric interactions between nematic alignment and flow and avoids spurious instabilities in the absence of an active inputs.
\end{abstract}

 \maketitle

\section{Why nematodynamic theory needs a renewal} 

The renewed interest in dynamics of nematic fluids has arisen owing to its wide applications in studies of active matter. Nematic order is commonly encountered in biological tissues \cite{saw18,silberzan,salbr23}, cells \cite{yo19,cell23}, and bacterial swarms \cite{beer,beer2} colonies \cite{bcol}, and biofilms \cite{bfil}. In all these systems, the constitutive elementary units are macroscopic, in contrast to molecular units of common nematic liquids. Nevertheless, theories of active nematic are commonly based on the established nematodynamic theory supplemented by a phenomenological active input.

The established theories aspire to be universal: they do not assume any specific mechanism of interactions between rearrangements of nematic order and flow, but derive dynamics close to equilibrium by assigning general linear relationships between thermodynamic forces and fluxes and establish relations between their coefficients via Onsager's reciprocal relations. The original Ericksen--Leslie (EL) approach \cite{erk,ls,dg} employing the nematic director as the order parameter, has been later extended to allow for a variable modulus through the use of the tensor order parameter \cite{eb94, qs98}, but all these theories, briefly reviewed below, rely on near-equilibrium relations between thermodynamic forces and fluxes. There are general objections to this approach \cite{bp} stating that a combination of variables even and odd  under time reversal  is \emph{physically unsound}, and cannot ensure evolution to thermodynamic equilibrium, but they were generally ignored, even after it has been shown by straightforward computation \cite{me22,mebook} that instability may arise under certain conditions in nematodynamic equations derived in this way even in the absence of an active input, which is ostensibly forbidden by Onsager's relations.

\section{The Lagrangian}

\subsection{From director to tensor order parameter}

The classical director-based description is deficient even in ``dry'' problems lacking hydrodynamic or mechanical interactions. The director $\mathbf{n}$ is formally a unit vector, but, unlike proper vectors, it is supposed to be symmetric to rotation by $\pi$, and therefore any energy expression should include an even power power of $\mathbf{n}$. The deviation of a static nematic medium from equilibrium maybe caused only by changes in orientation expressed by differential terms dependent on distortions of perfect alignment. The classical lowest-order expression for the energy density of a uniaxial nematic in 3D, the simplest form of the Landau--de Gennes Lagrangian going back to Frank\cite{fr} is 
\begin{equation} 
 \mathcal{L} =\tf12\left[  K_1( \mathrm{div}\, \mathbf{n})^2 + K_2 ( \mathbf{n} \cdot \mathrm{curl}\, \mathbf{n})^2 + K_3 ( \mathbf{n} \times \mathrm{curl}\, \mathbf{n})^2  \right].
 \label{elastK} \end{equation}
The three terms in this expression correspond to the energies of splay, bend, and twist distortion of nematic alignment. They include all possible quadratic combinations of the director with its divergence and curl invariant to rotations by $\pi$. The coefficients $K_i$, called Frank constants, are the respective elasticities. 

The dynamic equation that defines evolution of the nematic director in a quiescent medium, obtained by varying the Lagrangian \eqref{elastK}, is $\mathcal{D}_t \mathbf{n}_j=\Gamma\mathbf{h}_j$, where $\mathcal{D}_t=\partial_t + \mathbf{v}\cdot \nabla $ denotes the substantial derivative accounting for advection with flow velocity $\mathbf{v}$, $\Gamma$ is a mobility coefficient, commonly assumed to be a scalar, and $\mathbf{h}$ is the \emph{molecular field} characterizing orientational distortions: 
\begin{equation} 
h_i =\partial \mathcal{L}/\partial n_i + \partial_j \pi_{ji}, \qquad   \pi_{ji}= \partial\mathcal{L} /( \partial_jn_i).
 \label{hfield} \end{equation}
Summation over repeated indices is implied throughout. Generally, $\mathbf{n\cdot h}\neq 0$, so that the normalization of $\mathbf{n}$ is not preserved, but is sometimes sustained, without any physical justifications, by adding to \eqref{elastK} a Lagrange multiplier. Of course, the normalization $|\mathbf{n}|=1$ cannot be preserved in real nematic textures; for example, $\mathbf{n}$ vanishes or becomes indefinite in topological defects. Nevertheless, this approach is so far not abandoned, and the most commonly read and cited textbook on nematodynamics \cite{dg} does not go beyond the director-based description.

Realistic theories with variable modulus are commonly based on the tensor representation \cite{eb94,qs98,kl}. The tensor order parameter, commonly expressed for uniaxial nematics in $d$ dimensions by the symmetric traceless tensor $Q_{ij}=\varrho (n_in_j-\delta_{ij}/d) $ with the modulus $\varrho$, dependent on  coupled inclination angles, and therefore invariant to rotation by $\pi$.  The tensor description works perfectly in ``dry'' nematodynamics, although there are some problems with retaining all Frank constants while keeping the lowest-order Lagrangian \cite{mebook}. The ``wet'' problem is not as well settled. There are two distinct versions \cite{kl}, and both have been questioned \cite{bp} on the basis of an improper use of Onsager's reciprocity relations inherited from the original director-based description due to combining fields with opposite time-reversal symmetry.

\subsection{Vector-based description}

The progress from director-based to tensor-based theory skipped an intermediate possibility -- vector-based description, but this author \cite{me22, mebook} explored this option in 2D. Two-dimensional patterns are commonly observed in thin layers with tangential alignment on containing walls, as well as in thin elastic sheets and cellular layers, and computational models are commonly restricted to 2D, as 3D simulations are too cumbersome and hard to present on a 2D screen or paper sheet. In 2D, it is sufficient to characterize the nematic alignment by the vector order parameter $\mathbf{q}$ with the components
\begin{equation} 
q_1=\varrho^{1/2} (n_1^2-n_2^2)= \cos 2\theta, \;\;  q_2=2 n_1n_2 =\sin 2\theta.
\label{qn}  \end{equation} 
By construction, $|\mathbf{q}|=1$, and $\varrho$ is the modulus. Similar to the 2D tensor $\mathbf{Q}$, it is invariant to rotation by $\pi$, and  $\mathbf{Q}$ can be presented as a merger of $\mathbf{q}$ and its rotation by $\pi/2$, $\mathbf{q}^*$ with the components $q_2, \, -q_1$, leading to $Q_{11}=-Q_{22}=q_1, \, Q_{12}=Q_{21}=q_2$.  

For a uniaxial nematic, this representation be extended to 3D ensuring, as in 2D, the invariance to rotation by $\pi$. The expression constructed in this way, $\mathbf{q}_i =\tf13\varepsilon_{ijk}\mathbf{n}_j \mathbf{n}_k$ (where $\varepsilon_{ijk}$ is the 3D Levi-Civita antisymmetric symbol), can be specialized in different ways invariant to rotation by $\pi$. The form most suited to the above 2D expression is constructed by defining any $q_i$ as a 2D vector with the components  $\cos 2\theta_{jk},\, \sin 2\theta_{jk}$, where $\theta_{jk}$ is an an angle in the plane normal to $\mathbf{q}_i$.  

The 3D vector-based analog of the Lagrangian \eqref{elastK} can be also constructed by imitating the latter in a chosen way while replacing $\mathbf{n}_i$ by $\mathbf{q}_i$ and adding the algebraic term defining the modulus:
\begin{equation} 
\mathcal{L}=\alpha -\varrho^2+\mathcal{L}_q, 
\label{Lrr}  \end{equation} 
where $\alpha$ is the bifurcation parameter defining the transition to an anisotropic state at $\alpha >0$. 

The lowest-order distortion-dependent part of the Lagrangian $\mathcal{L}_q$ representing the elastic energy per unit volume can be also constructed in alternative ways, out of which we choose the one most similar by its structure to its director-based analog \eqref{elastK}:
\begin{align} 
\mathcal{L}_q&= \tf12  \Sigma_i \left\{K_1[\mathrm{div}\, \mathbf{q}_i]^2
 + K_2 [\mathbf{q}_i \cdot (\mathbf{q}_j \times \mathbf{q}_k)]^2 \right. \cr
&+\left. K_3 [\mathbf{q}_i \times  (\mathbf{q}_j \times \mathbf{q}_k)]^2\right\} .
\label{elastP}  \end{align} 
The first (splay) term is unique, while the choice of $\mathbf{q}_i$ in the bend and twist terms is arbitrary. This construction can be extended to biaxial nematics by supplementing the three parameters defining the inclination of the main axis by a 2D vector characterizing the projection of a secondary director on the plane normal to the primary one. 

In the framework of the``dry'' nematodynamics, the usage of the vector order parameter enables applying the classical EL formalism to systems with a variable modulus with minimal adjustments. However, it doesn't help a harsher problem: the inadequacy of applying Onsager's relations for derivation of a complete set of nematodynamic equations coupling nematic reorientation and flow. This procedure involves arbitrary tumbling parameters, and instabilities, forbidden in passive systems have been observed within some range of the latter's values \cite{me22, mebook}, suggesting that substantial changes of the approach are necessary. 

\section{Vector Nematodynamics}

\subsection{Interactions induced by rotational symmetry}

The primary cause of the change of nematic orientation is the tendency to ordering, characterized by the molecular field defined in the vector-based theory, similar to Eq.~\eqref{hfield} but extended in 3D to the triple order parameter set in boldface, as
\begin{equation} 
\mathbf{h}_i =\partial\mathcal{L}/\partial \mathbf{q}_i + \partial_j \bs{\pi}_{ji}, \quad   \bs{\pi}_{ji}= \partial\mathcal{L}_q /( \partial_j\mathbf{q}_i).
 \label{qfield} \end{equation}

A complementary course of reorientation, overlooked by extant theories, is rotation of nematic alignment by local vorticity. It follows from the \emph{local} symmetry between simultaneous rotation of nematic alignment and flow. Since $\mathbf{q}_i$ rotate with twice director's speed, the vector-based energy-conserving relation is written as 
\begin{equation}
\partial_t \mathbf{q}_i= 2A^-_{ij}\mathbf{q}_j, \quad \partial_t x_i=A^-_{ij}x_j,  
\label{rotom}\end{equation}
where $A^-_{ij}$ is the antisymmetric part of the rate of strain tensor $A^\pm_{ij}=\tf12 (\partial_iv_j \pm \partial_jv_i) $. In standard theories, this symmetry is viewed as \emph{global}, which conceals its role in the exchange between elastic and hydrodynamic energy. The vector 
\begin{equation}
\bs{\varpi}_i=2\mathbf{A}^- \cdot \mathbf{q}_i
\label{varpi}\end{equation}
represents the rate of rotation of the order parameter by fluid flow, lowering the latter's energy, as distinguished from rotation driven by lowering the elastic energy. Thus, the total rotation is computed as 
\begin{equation}
\partial_t \mathbf{q}_i= \Gamma\mathbf{h}_i+\bs{\varpi}_i,  
\label{dq}\end{equation}
while the vectors $\bs{\varpi}_i$ enter the hydrodynamic energy balance, as discussed below. 

A so far overlooked aspect of  the rotational symmetry is its role in nematodynamic energy and momentum balance.  While rotation driven by reducing nematic energy generates the distortion stress $\bs{\sigma}^\mathrm{d}_{ij}=-\bs{\pi}_{kj}\partial_i q_k$, the symmetry-driven rotation is purely hydrodynamic and induces the antisymmetric viscous stress, overlooked both by director- and tensor-based theories and taken in account so far only in this author's study of a specific model \cite{me24}.
 
 \subsection{Energy exchange}

Elasto-hydrodynamic interactions involve exchange between the elastic energy $\mathcal{E}= \int \mathcal{L}\,\D\mathbf{x}$ and flow energy $\mathcal{F}=\tf12\int \rho |\mathbf{v}|^2\D\mathbf{x}$. The latter's change with time depends on acceleration of the fluid determined by the generalized Navier--Stokes (NS) equation $\rho \mathcal{D}_t v_i = f_i$, where $\rho$ is the fluid's density and $\mathbf{f}(\textbf{x})$ is the total force acting upon the fluid. In the standard hydrodynamics of isotropic fluids, this force includes the pressure gradient $\nabla p$ and the symmetric viscous stress $\bs{\sigma}^+$, dependent linearly on the symmetric rate of strain $\mathbf{A}^+$.  In an oriented fluid, this relation is generally anisotropic, $\sigma^+_{ij} = \eta_{ijkl}^+A^+_{kl}$, with viscosities $\eta^+_{ijkl}$ expressed as fourth order tensors built up of combinations of $\mathbf{q}_i$ (or $n_i$ in the EL theory) with empirical coefficients. 

In extant nematodynamic theories based on the entropy, rather than energy, balance, the NS equation is complemented by the gradient of the distortion stress stemming from the reduction of elastic energy. It plays a secondary role in the analysis, since it is quadratic in small perturbations \cite{mebook}, but its inclusion is not justified in the framework of the present approach, since the energy released due to nematic ordering may be dissipated as heat rather than affecting directed fluid motion. 

In the framework of the present theory, the hydrodynamic force related to the change of nematic orientation originates in the flow energy driving the rotation of the order parameter, as given by the first relation \eqref{rotom}. Rotation costs energy, and is counteracted by viscosity, generally anisotropic, generating the antisymmetric stress $\sigma^-_{ij} =-2\eta_{ijk}^-A^-_{kl}\mathbf{q}_l$ with antisymmetric (odd) viscosity $\eta_{ijk}$ dependent on local orientation. This relation, alongside the velocity-dependent Eq.~\eqref{dq}, carries the connection between flow and elastic dynamics. 

Neglecting viscous anisotropy, the applicable hydrodynamic balance is expressed as
\begin{equation}
\rho\mathcal{D}_t v_i=\partial_i(\eta^+ A^+_{jl} -2 \eta^-\varepsilon_{jk} A^-_{kl})q_l -\partial_i p. 
\label{NS}\end{equation}

 The total energy or enthalpy $\mathcal{H}=\mathcal{E}+\mathcal{F} -\int p\, \D\mathbf{x}$
 must decrease in a system evolving to equilibrium. Unlike all extant theories, this criterion does not rely on Onsager's relations and can be applied as well to active systems driven far from equilibrium by added active stress. Of course, implementing this criterion requires finding both the order parameter and flow velocities by solving the dynamic equations for the order parameter and velocity, but stability of an ordered state to weak perturbations can be established analytically as follows.

 \section{Stability of a Quiescent State}

As a basic example of application of stability analysis, consider perturbations of an ordered state in an unbounded domain. The simplest case is a 2D problem with $\mathbf{x}=\{x_1=x_\|,\, x_2=x_\perp$ and the base state $q_1=q_\|=1, \, q_2=q_\perp=0$, realizable in an adsorbed layer or a thin sheet with planar orientation imposed by boundary conditions on confining walls. This approach can be extended  to 3D in the cylindrical geometry, when rotational velocity is irrelevant for perturbations of the alignment along the symmetry axis. 

The $\mathcal{O}(\epsilon)$ deviations from the base state and flow velocity are expanded in the Fourier series 
\begin{equation}
q_i = \epsilon \int\widehat{q}_i(\mathbf{ k}) \E^{\I\mathbf{ k\cdot x}}\D\mathbf{ k}, \qquad
\widetilde{v}_i = \epsilon\int \widehat{v}_i(\mathbf{ k}) \E^{\I\mathbf{ k\cdot x}}\D\mathbf{ k}.
\label{lin}\end{equation}
It is advantageous to express the axial and normal velocities through the stream function $\Psi$ as $\widetilde{v}_i =\varepsilon_{ij}\partial_j \Psi$. The stream function is also expanded in the Fourier series $\widehat{\Psi}(\mathbf{ k})\E^{\I\mathbf{ k\cdot x}}$. After expressing the components of the wave vector as $k_1=\kappa \cos \theta, \,k_2=\kappa \sin \theta$, the Fourier components of the symmetric and antisymmetric strains ${A}^\pm$ take the form
\begin{align}
\mathbf{A}^+ &=- \epsilon\kappa^2 
\begin{pmatrix}
    \sin^2 \theta  & \tf12 \cos 2 \theta \\
      \tf12 \cos 2 \theta&  - \sin^2 \theta
\end{pmatrix}\widehat \Psi,  \cr
\quad \mathbf{A}^- &=- \epsilon\kappa^2 
\begin{pmatrix}
    0  & \tf12\\   - \tf12  &  0 \end{pmatrix} \widehat\Psi  
    = -\tf12 \epsilon\kappa^2\bs{\varepsilon}\widehat\Psi. 
\label{ao}\end{align}

In an incompressible fluid, pressure can be eliminated by applying the curl $\varepsilon_{ij}\partial_j$ to the hydrodynamic equation \eqref{NS}. Using the identity $\varepsilon_{ij}\varepsilon_{jk}=\delta_{ik}$ to simplify the antisymmetric term, we obtain 
\begin{equation}
\varepsilon_{ij}\rho\mathcal{D}_t\partial_jv_i=\eta^+\varepsilon_{ij}\partial_i\partial_l A^+_{jl} -2 \eta^-\partial_i A^-_{il}q_l. 
\label{NS2}\end{equation}
In the lowest $\mathcal{O}(\epsilon)$ order, $q_l$ has a single nonzero component $q_1=1$ The resulting $\mathcal{O}(\epsilon)$ Fourier transform of Eq.~\eqref{NS2} is
\begin{equation}
\rho\mathcal{D}_t \widehat{\Psi} 
= - (\tf12 \eta^+ \cos^2 2\theta+\I \eta^- \cos\theta)\widehat{\Psi}. 
\label{NSF}\end{equation}
The real part of this expression testifies decay of convective fluctuations due to the standard symmetric viscosity. The imaginary part, originating in interaction between fluctuations of nematic alignment and flow, causes the decay to be oscillatory, so that, e.g., fluctuations normal to the original alignment generate fluctuations in the parallel direction before both are decaying. 

The linearized molecular field \eqref{qfield} following from the Lagrangian \eqref{elastP}, complemented by the algebraic part, reduced after setting $\alpha=1$ to a single component $\widetilde{h}_1 =-2 \widetilde{q}_1$, is stabilizing. The added convective term \eqref{varpi}, linearized using Eq.~\eqref{ao}, contributes $\mathcal{O}(\epsilon)$ perturbations only to the equation of $q_2$, adding to its Fourier expansion the term $-\kappa^2\widehat{\Psi}$, which decays together with $\mathcal{O}(\epsilon)$ convection. Thus, relaxation to the ordered state is ensured in a passive system as it must, unlike spurious instabilities in theories based on the Onsager's relations \cite{me22, mebook}.

A fully 3D analysis is facilitated in the case of an incompressible fluid by expressing its velocity  through the \emph{vector} analog of the stream function as $v_i=\varepsilon_{ijk}\partial_j\Psi_k$ with $\nabla \cdot \bs{\Psi}=0$ \cite{psi}. With $\Psi \sim{O}(\epsilon)$, the symmetric and antisymmetric strains are presented as 
\begin{equation}
A^\pm_{il}= \epsilon(\varepsilon_{ijk}\partial_j\partial_i \pm\varepsilon_{ljk}\partial_j\partial_l)\Psi_k. 
\label{ao3}\end{equation}
Also in this case, pressure is eliminated by applying to Eq.~\eqref{NS} the curl $\varepsilon_{ijl}\partial_j$, but this device is not so useful, as it does not reduce the number of equations. Also in this case, the leading-order system depends only on the unperturbed orientation and lacks potentially destabilizing terms.   
 
\section{Discussion} 

The suggested mechanism of interactions between changes of nematic alignment and flow is not dependent on proximity to equilibrium and follows naturally from the symmetry between rotation of nematic alignment and flow vorticity, which, unlike its common treatment, should induce antisymmetric stress. 

The established way to derive alignment-flow interactions through Onsager's reciprocal relations, aspiring to be model-independent, turns out to be faulty, as it allows for spurious instabilities, which can be traced to an arbitrary parameter emerging in the course of derivations, which adds orientation-dependent terms to the hydrodynamic balance. It is often called ``tumbling parameter'', although it is not formally connected with the actual tumbling phenomenon: rotation of nematic orientation in order to align with flow in a specific way. It is unlikely to be significant on the molecular scale, but may be important in orientable media with macroscopic basic elements common to active matter. However,   tumbling rotation may induce only an antisymmetric stress, similar to but weaker than that described here and cannot be destabilizing.

The wide and  long-time application of the established nematodynamic theory may invalidate the results of certain derivations and simulations on various scales, from molecular to macroscopic, so that they would need revision. Flow-alignment interactions and the way they are affected by activity are likely to be specific in different applications, especially biologically related, and require further deep insights.

\end{document}